\documentclass{aa}

\usepackage{graphics,latexsym,psfig}

\def\etal{{et\,al.}}

\def\WD{white dwarf}
\def\msun{M$_{\odot}$}

\def\mdot{$\dot M$}

\def\amin{\ifmmode ^{\prime}\else$^{\prime}$\fi}
\def\asec{\ifmmode ^{\prime\prime}\else$^{\prime\prime}$\fi}
\def\fm{\hbox{$.\!\!^{\rm m}$}}            

\newbox\grsign \setbox\grsign=\hbox{$>$}
\newdimen\grdimen \grdimen=\ht\grsign
\newbox\laxbox \newbox\gaxbox
\setbox\gaxbox=\hbox{\raise.5ex\hbox{$>$}\llap
     {\lower.5ex\hbox{$\sim$}}}\ht1=\grdimen\dp1=0pt
\setbox\laxbox=\hbox{\raise.5ex\hbox{$<$}\llap
     {\lower.5ex\hbox{$\sim$}}}\ht2=\grdimen\dp2=0pt
\def\gax{\mathrel{\copy\gaxbox}}
\def\lax{\mathrel{\copy\laxbox}}
\def\rxj0513{RX\,J0513.9--6951}
\def\1e{1E 0035.4--7230}
\def\chan{{\it Chandra}}

\begin{document}

\title{X-Ray Off States and Optical Variability in CAL 83}

 \author{J. Greiner\inst{1,2} \and R. Di\,Stefano\inst{3, 4}}

 \offprints{J. Greiner, jgreiner@mpe.mpg.de}

 \institute{Astrophysical Institute
      Potsdam, An der Sternwarte 16, 14482 Potsdam, Germany
    \and
      Max-Planck-Institute for extraterrestrial Physics, 85741 Garching, 
      Germany
    \and
      Harvard-Smithsonian Center for Astrophysics, Cambridge, MA 02138
    \and
      Department of Physics and Astronomy, Tufts University, Medford, MA 02155
     }

\authorrunning{Greiner \& Di\,Stefano}

   \date{Received 30 July 2001 / Accepted ?? March 2002 }
   \abstract{
CAL 83 was one of the first supersoft X-ray binaries (SSBs) to be discovered
and is considered to be the prototype of its class. In $15$ X-ray
observations between 1983--1997 it was observed to have nearly constant
X-ray luminosity and temperature, with the exception of one off-state
in 1996 (Kahabka {\it et al.}\ 1996). We report on a second X-ray
off-state, discovered with a \chan\ ACIS-S observation in November 1999.
We consider the long-term X-ray and MACHO optical light curves. 
We find that, during more than $7$ years of monitoring by the MACHO team,
 CAL 83 has exhibited  distinct and well-defined
 low, intermediate, and high optical states.
Transitions between states are not accompanied by color variations.
We also find that both X-ray off states were observed during
optical high states and were followed by optical low states within
$\sim 50$ days. We discuss possible explanations for the observed 
optical and X-ray variations.
While photospheric adjustments might account for the variations in
soft X-ray flux, optical variations can be explained only by invoking
changes in the accretion disk, which is the primary source of
optical radiation.
      }

\maketitle

\section{Introduction}

Supersoft X-ray sources are sources whose
radiation is almost  completely emitted in the energy band below 0.5 keV and
whose bolometric luminosity is 10$^{36-38}$ erg s$^{-1}$.
Some SSSs are hot white dwarfs (WDs) in a post-nova phase, and some
are WD or pre-WD central stars of planetary nebulae. An important subclass
of SSSs are supersoft X-ray binaries (SSBs). 
The first two SSBs to be discovered, CAL 83 and CAL 87, were observed
with {\it Einstein} (Long \etal\ 1981), and were found 
to have optical properties (luminosity and periodicity) 
very similar to the properties of low-mass X-ray binaries
(Crampton \etal\ 1987; Pakull \etal\ 1987, 1988).
ROSAT observations revealed many more of these sources, and established
supersoft X-ray binaries as a class (see Greiner 1996 and references therein).
Several different observational facts suggest that
SSBs contain white dwarfs and that 
many of these are burning hydrogen in a shell, with a very thin and
hot atmosphere on top. The widely accepted
scenario for these so-called close-binary supersoft sources
(Di Stefano \& Nelson 1996), which have orbital periods of tens of hours, 
is based on a companion which may be slightly evolved and/or
more massive than the white dwarf. These conditions allow  
mass transfer to occur at the high rates required
to burn the accreted hydrogen in a quasi-steady manner near the white dwarf's 
surface (van den Heuvel \etal\ 1992; Rappaport \etal\ 1994).
SSBs are known to be variable on time scales 
from hours to years (see Greiner 1995 for a review; 
or Kahabka 1996; Krautter \etal\ 1996; Greiner \etal\ 1996;
Reinsch \etal\ 2000).

CAL 83 is the prototypical supersoft X-ray binary (SSB), and among the 
brightest (in terms of count rate in X-ray detectors) at X-ray wavelengths.
CAL 83 was long thought to be an X-ray persistent SSB, until Kahabka \etal\
(1996) discovered an X-ray off state in April 1996. 
Observations taken $\sim 20$ days prior to and $\sim 100$ days after
the off-state, found CAL 83 to exhibit its normal X-ray brightness,
thus constraining the duration of the off-state to less than 120 days
(Kahabka 1998).

IUE and HST ultraviolet observations showed that the UV flux is variable 
by more than a factor of two
(Crampton \etal\ 1987; G\"ansicke \etal\ 1998). During the UV-bright state
the optical emission was also brighter (Bianchi \& Pakull 1988).
Crampton \etal\ (1987) claimed that during the UV-bright state
the IUE-spectrum was steeper, thus suggesting a hotter source,
but a re-analysis did not reproduce this finding (G\"ansicke 2002,
priv. comm.).
Besides the orbital, sinusoidal variation of 1.0436$\pm$0.0044 days
(Smale \etal\ 1988), CAL 83 has also shown seemingly irregular, optical 
variability on a longer time scale (Pakull \etal\ 1985; Crampton \etal\ 1987;  
Bianchi \& Pakull 1988).

 \begin{figure*}
   \vbox{\psfig{figure=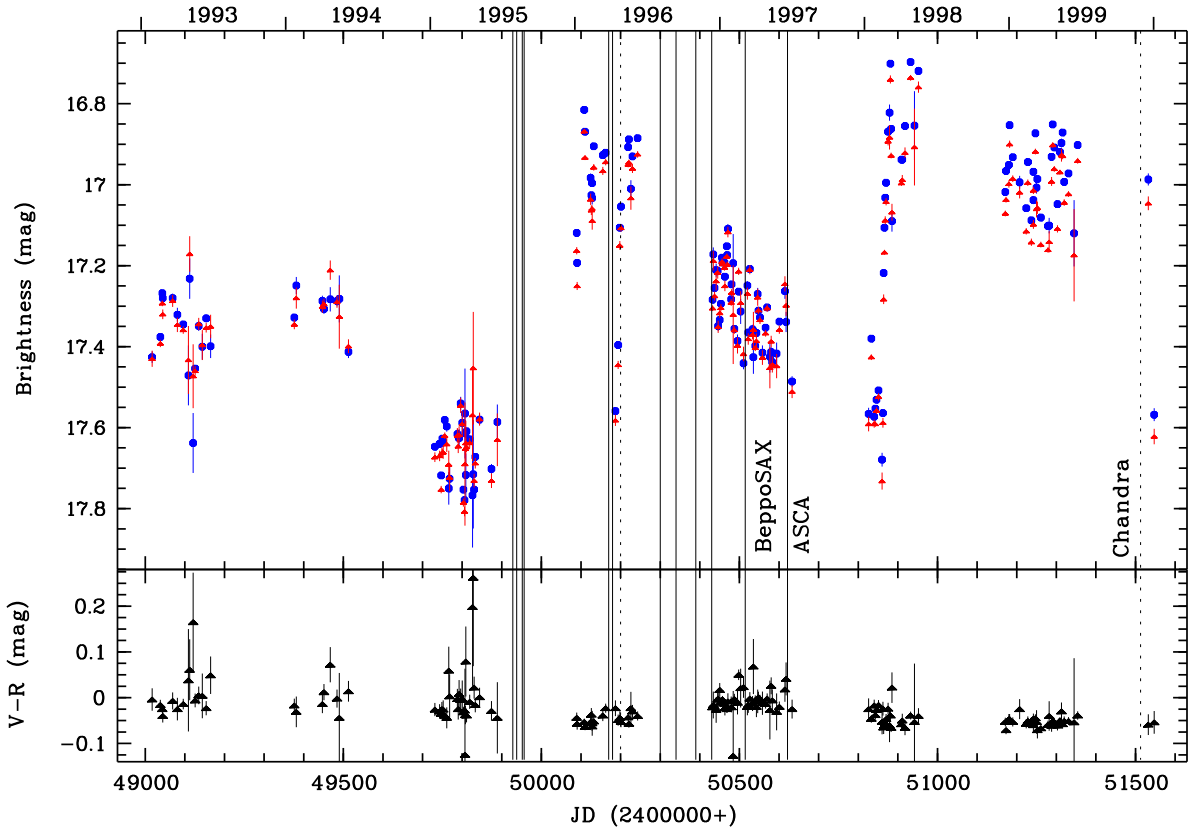,width=16.cm,angle=270,%
          bbllx=2.9cm,bblly=2.6cm,bburx=19.2cm,bbury=25.7cm,clip=}}
   \vspace*{-0.2cm}
   \caption[twocol]{MACHO light curve of CAL 83 based solely on the 
      two-color calibrated data (see text for more details).
      Visual-filter ($V$ band)  observations are marked by filled circles, 
      while red-filter ($R$) are marked by  triangles. The lower panel
      shows the $V-R$ color. Vertical lines denote times of X-ray observations.
      With the exception of observations with {\it Beppo}SAX and {\it Chandra}
      (individually labeled) all of these have been performed with ROSAT. 
      The full lines indicate normal X-ray ``on''-state, while dotted lines 
      mark the two only X-ray ``off''-states. 
   \label{twocol}}
 \end{figure*}

CAL 83 happened to be in a field regularly monitored by the MACHO team
for more than 7 years. Based on a $\sim$150 day portion of the MACHO
observations centered around the X-ray off-state in 1996 (modified
Julian Date MJD $\sim$ 50200), Alcock \etal\ (1997) noticed
an optical minimum which preceded the X-ray off-state by $\sim$10--15 days.
Alcock \etal\ (1997) concluded that this sequence of events is similar
to the decline phase of a nova.
Alternatively, Kahabka (1998) suggests that variations in the temperature
of the white dwarf envelope due to expansion/contraction are the cause for
the observed X-ray variability, similar to RX J0513.9--6951 
(Pakull \etal\ 1993; Reinsch \etal\
1996; Southwell \etal\ 1996) and AG Dra (Greiner \etal\ 1997).

In this paper we add two elements that lead to an alternative interpretation.
First, we have used {\it Chandra} to observe a second X-ray off-state.
In contrast to the 1996 X-ray off-state, this one took place during
a well-sampled optical high state and appears to have {\it not}
been preceded by a recent optical low state. Second, we now have access to
more than $7$ years of data from the optical monitoring of the LMC by
the MACHO team. We therefore
combine the MACHO data with all available X-ray observations. We find
that both X-ray off states occurred during optical high states, and that
both were {\it followed} by an optical low state.
We discuss the synthesis of all of the optical and X-ray data on CAL 83
and possible interpretations of the patterns therein.

\section{Observations and Results}

\subsection{Optical Observations by the MACHO team}

The MACHO Project (Alcock \etal\ 1995) was a microlensing survey
that monitored the brightnesses of
60 million stars in the Large Magellanic Cloud, Small Magellanic Cloud,
and Galactic bulge between 1992 and end of 1999. It used the
1.27 m Great Melbourne Telescope located at the Mount Stromlo Observatory
in Australia. Two filters were used: a visual filter (4500--6300 \AA) and
a red filter (6300--7600 \AA), the magnitudes of which are transformed
to the standard Kron-Cousins V and R system, respectively.
At both the red and visual foci, a mosaic
of four 2048 $\times$ 2048 Loral CCDs are mounted.

The optical data were kindly provided by the MACHO team (courtesy A. Becker).
A total of 514 observations of CAL 83 have been performed, 168 in the red
and 346 in the visual filter, respectively.

Because of the rather complicated calibration issues (Alcock \etal\ 1999)
involved in the transformation of the MACHO instrumental magnitudes into
the standard Kron-Cousins V and R system, only those measurements can be
unambiguously calibrated for which (nearly-) simultaneous red- and 
visual-filter
observations are available. For CAL 83 this is the case for 130 observations
(called two-color data in the following), while the remainder are one-color
measurements. The remaining calibration zero-point uncertainty for the
MACHO photometry database is estimated to be $\pm$0.10 mag in $V$ or $R$,
 and $\pm$0.04 mag in $R$-$V$.

\begin{figure}
     \vbox{\psfig{figure=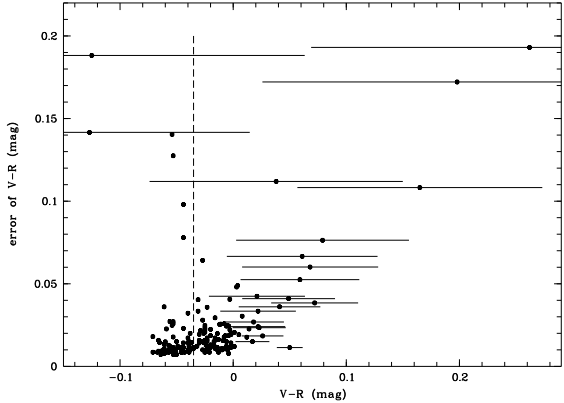,width=8.3cm,angle=270}}
   \vspace*{-0.3cm}
   \caption[rmv]{Error of the color 'B'-'R' = $V$-$R$ plotted
          over the color. Nearly all data points with more than 0.05 mag
          deviation from the median (dashed lines) have errors which
          make them compatible with the median.
     \label{RmV}}
   \end{figure}

The light curve of CAL 83 based on the two-color measurements is shown in
Fig. \ref{twocol}, including the $R$-$V$ color.
It can be seen that only marginal color changes
are associated with
the large intensity changes (Fig. \ref{twocol}). 
One might be tempted to ``see'' the color being bluer during optical
high-intensity states. However, nearly all of the data
points with more than 0.05 mag color variation relative to the mean
value also have rather large errors (bottom panel of Fig. \ref{twocol},
Fig. \ref{RmV}). From this we conclude that the $R$-$V$ color is basically 
constant, formally $\Delta$($V-R$) = 0.03$\pm$0.05 mag,
throughout the $>$1 mag intensity changes.

  \begin{figure}
     \vbox{\psfig{figure=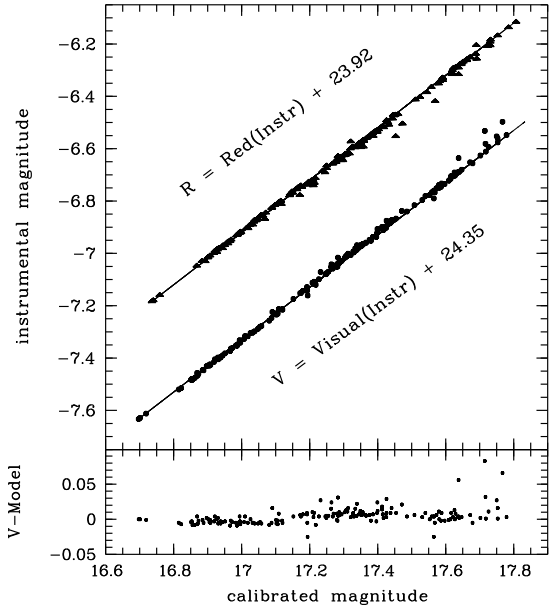,width=8.3cm,angle=270,%
          bbllx=1.9cm,bblly=2.2cm,bburx=19.3cm,bbury=19.8cm,clip=}}
   \vspace*{-0.3cm}
   \caption[cross]{Correlation of the calibrated versus instrumental
           magnitudes in the two-color calibrated data (top panel). With few
           exceptions, the linear correlations as labeled are accurate
           to better than $\pm$0.05 mag,
           though a non-linear term could be present (shown in the
            lower panel for the V band data). We have used the linear term
           for a ``generic'' calibration of the one-color data 
           (Fig. \ref{macho}).
     \label{cross}}
     \end{figure}

In the case that the color of an object is constant, a ``generic''
calibration of the one-color data to a standard system can be made.
As shown above, this is justified for CAL 83.
We determine a mean offset between instrumental and calibrated magnitudes
from the two-color data set of 24.35 mag in the `B' band, and 23.92 mag 
in the `R' band (Fig. \ref{cross}). 
There could be a small ($<$0.02 mag) systematic non-linear term in the 
V band data, with the instrumental magnitude being lower (higher)
at the bright (faint) end of the calibrated magnitudes (see lower panel
of Fig. \ref{cross}), but this is smaller
than the quoted MACHO errors, so we ignore it in the following.
We note, however, that if this were true, it would reduce the 
$\Delta$($V-R$) even further.
These offsets have been applied (with the
above justification) to the one-color data, and the light curve using
the complete data set is shown in Fig. \ref{macho}.

Before turning to the obvious variability seen in the data,
it is worth mentioning that CAL 83 exhibits orbital variation of its
intensity of about $\pm$0.11 mag (Smale \etal\ 1988).
Since the MACHO observations were not done at identical phases, these orbital
variations have to be kept in mind when interpreting the light curve.

\subsection{X-ray observations}

We present the results of our recent {\it Chandra} observation 
and also summarize results of previous X-ray observations. 

\subsubsection{{\it Einstein}}

CAL 83 was the softest and brightest source in the LMC survey which 
comprised 97 sources (Long \etal\ 1981). A comprehensive analysis of the
six {\it Einstein} observations of CAL 83 is given by Brown \etal\ (1994).
For the two long-exposure IPC observations the spectral fits result
in $kT \sim 8-80$ eV depending on the adopted absorbing column.
The original count rate values of Brown \etal\ (1994)
imply a range of 0.006$\pm$0.002 to 0.123$\pm$0.016 cts/s.
However, the observations were done with CAL 83 located near the
edge of the field of view of the detector, with off-axis angles in the 
25\amin -- 35\amin\ range, thus introducing possible systematic uncertainties;
furthermore, Kahabka (1998) mentions a private communication by E. Gotthelf
that the implied variability could be entirely due to instrumental effects,
like vignetting correction or obstruction by detector ribs. 
With the permission of E. Gotthelf we publish his results (which take
into account these corrections) in the appendix,
which indeed show CAL 83 to be constant during all IPC observations.

  \begin{figure*}
    \vbox{\psfig{figure=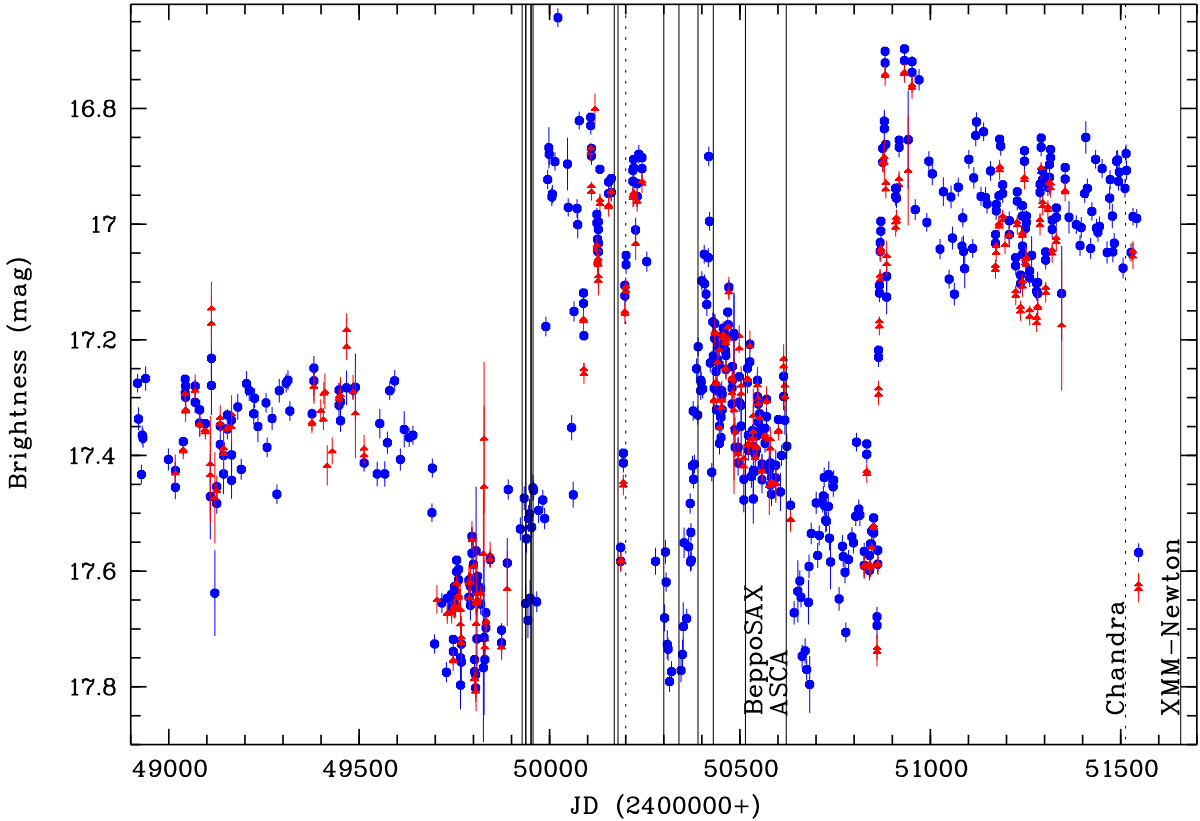,width=16.cm,angle=270,%
          bbllx=2.7cm,bblly=2.cm,bburx=19.4cm,bbury=26.0cm,clip=}}
   \vspace*{-0.2cm}
   \caption[macho]{MACHO light curve of CAL 83 showing also the
         one-color data after the ``generic'' correction. Times of
           X-ray measurements are again overplotted (see Fig. \ref{twocol}).
   \label{macho}}
   \end{figure*}

\subsubsection{EXOSAT}

CAL 83 was observed by EXOSAT on 1985 December 18 over a period of about 16 hrs
(Crampton \etal\ 1987). Most of the CMA observations were made with the 
Lexan filter (0.05--2 keV), yielding a mean count rate of 0.0422$\pm$0.0014
cts/s, but some data were also obtained with the Al/P filter 
(0.0204$\pm$0.0032 cts/s). Since the Boron filter observation only 
provided an upper limit, not much spectral information can be deduced
apart from the fact that the spectrum must be soft.
Contemporaneous optical spectroscopy shows no peculiarities with respect
to earlier spectroscopic data, suggesting no outburst or unusual state.

\subsubsection{ROSAT}

CAL 83 was observed by ROSAT several times, first during the verification
phase in June 1990 (Greiner \etal\ 1991), and then during the 
ROSAT all-sky survey (1990/1991). These observations show CAL 83 at
$\sim$1 cts/s in the PSPC. Spectral fitting using a blackbody model
gives best-fit values of
25--45 eV and a luminosity around 10$^{38}$ erg s$^{-1}$ (Greiner \etal\ 1991).
The small variations seen in count rate and spectral shape are consistent 
with being due to varying response at different off-axis angles. 
All ROSAT pointed observations after the all-sky survey have been 
obtained with the HRI and were summarized and extensively discussed
by Kahabka (1998). Since the MACHO data only cover the time since 1992,
we just display the times of pointed ROSAT observations as given in Tab. 1
of Kahabka (1998). The observed count rates, except for the one observation
during the April 1996 off-state, show a scatter of less than 20\% around
the mean of $\sim$0.2 HRI cts/s. The X-ray off-state observation 
gives an 2$\sigma$ upper limit of 4.8$\times$10$^{-3}$ cts/s,
which could be modeled by a temperature decrease from 39 eV to 27 eV
(Kahabka 1998), though the exact range depends on the absolute temperatures
since the count rate is a strong non-linear function with
decreasing temperature.

\subsubsection{{\it Beppo}SAX}

Among several other SSBs, CAL 83 was observed with {\it Beppo}SAX
on March 7/8 1997 for a total of 36 ksec in the Low-Energy Concentrator 
Spectrometer (LECS).
No absorption edges have been found in this observation which could 
hint to a hot white dwarf atmosphere (though the low signal-to-noise
ratio allows no further conclusions). Blackbody and white dwarf atmosphere
models fit equally well, and result in a temperature of 
30--50 eV and a luminosity of 2--6$\times$10$^{37}$ erg s$^{-1}$ 
depending on which model is used and whether the absorbing column is
fixed to the HST-derived value or left free (Parmar \etal\ 1998),
consistent with the values derived from the ROSAT PSPC observations.

\subsubsection{ASCA}

An ASCA observation of CAL 83 was made on 21/22 June 1997, for a total
of 16 ksec. The X-ray intensity was as expected for CAL 83's normal X-ray on 
state, and even a combined fit with ROSAT data from June 1990 was possible
without any need for relative intensity adjustments (Dotani \etal\ 2000).
As with the {\it Beppo}SAX data, the higher spectral resolution data did not
require any absorption edges, and the derived effective temperature
(29 $\pm$8 eV) was identical to the ROSAT value (Dotani \etal\ 2000).

\subsubsection{\chan}

CAL 83 was observed within the Guaranteed-Time Observations programme 
with ACIS-S (PI: S. Murray) on 30 November 1999 for 2.11 ksec. 
Because of the better soft X-ray response, the backside-illuminated ACIS-S 
was chosen in the imaging observation. Since we expected about 2 cts/s,
the observation was performed in continuous clocking mode.
CAL 83 remained undetected, the 3$\sigma$ upper limit is 0.004 cts/s. 
Under the assumption of kT$\sim$40 eV 
(and an absorbing column of $N_{\rm H} = 6.5\times 10^{20}$ cm$^{-2}$;
G\"ansicke \etal\ 1998) this corresponds to a luminosity limit of
8$\times 10^{34}$ erg s$^{-1}$. If, on the contrary, we assume a constant 
bolometric luminosity of 4$\times 10^{37}$ erg s$^{-1}$,
the non-detection implies either a temperature limit of kT $\lax$ 15 eV
at the canonical absorbing column, or a limit 
on the absorbing column of $N_{\rm H} \gax\ 6\times 10^{21}$ cm$^{-2}$
at the canonical 40 eV effective temperature.

  \begin{figure*}
    \vbox{\psfig{figure=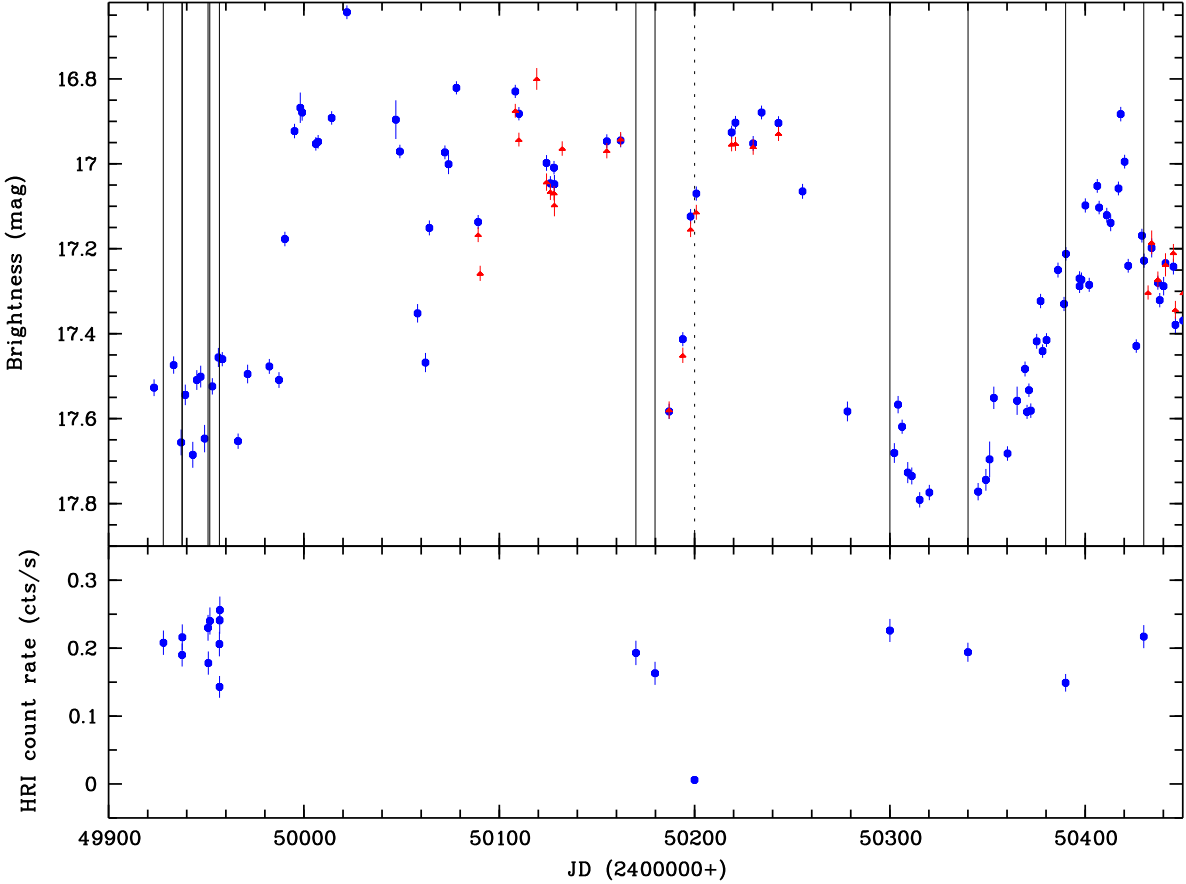,width=16.cm,angle=270,%
          bbllx=2.cm,bblly=2.8cm,bburx=19.3cm,bbury=25.8cm,clip=}}
   \vspace*{-0.2cm}
   \caption[macho]{Blown-up MACHO light curve of CAL 83 covering all
     the times of pointed ROSAT HRI observations. Times of
     X-ray measurements are again overplotted (see Fig. \ref{twocol}).
     The lower panel shows the ROSAT HRI count rates as published by
     Kahabka (1998).
   \label{hri}}
   \end{figure*}

CAL 83 was also observed with HRC-S+LETG (PI: Brinkman) for 52.3 ksec on 
29 November 1999. The non-detection in HRC-S (Murray, priv. comm.)
at zeroth order implies an  upper limit which is of the same order
of magnitude as the ACIS-S limit (the factor 25 in exposure time
is balanced by a factor 16--20 less efficiency of the grating).

\subsubsection{XMM-{\it Newton}}

On 23 April 2000, CAL 83 was observed with XMM-{\it Newton} for
a total of 45 ksec, aiming at a 
high-resolution X-ray spectrum using the Reflection Grating Spectrometer (RGS)
(Paerels \etal\ 2001).
While the spectrum shows a surprising wealth of details, the source intensity
is as expected from extrapolation of observed intensities with ROSAT and 
{\it Beppo}SAX (using PIMMS), 
thus indicating that by the date of this observation
CAL 83 was back to its normal X-ray on-state.

\section{Patterns in the Data}

\subsection{Pattern of Optical Variations} 

During more than 7 years of MACHO observations, CAL 83 has been observed 
to exhibit three apparently distinct optical states.
There is (1) a low state, during which the magnitude varies between
roughly 17.8 and 17.5 mag;
(2) an intermediate state, during which the magnitude varies between
roughly 17.5 and 17.1 mag;
and (3) a high state, during which the magnitude varies between
roughly 17.1 and 16.7 mag. 
The complete MACHO light curve is displayed in Fig. \ref{macho}; here we
mention some of its salient features.

(a) Although there are transitions between states, the states themselves
are well defined.
That is, the time spent continuously within any given state
is generally significantly longer than the time spent
making transitions into or
out of that state.

(b) In some portions of the optical light curve, variations about the 
mean magnitude within a state (low, intermediate, or high) seem
to be comparable to the $\sim \pm$ 0\fm11 mag orbital
variations (Smale \etal\ 1988), e.g. 
during the time periods (in modified Julian Date) 49000--49600, 
49700--49900, and 50400--50600.
These variations are, of course, occurring everywhere in the light curve,
but are difficult to recognize (without a period folding analysis)
during times of intrinsic variability.

(c) When the system makes a transition to a higher state, it tends 
to ``overshoot"; i.e., the luminosity soon after the transition is generally
significantly ($\sim 0.2$ mag) larger than the average luminosity of the
system during the rest of the time in that higher state.
Each of the  three cases of
transitions to a higher state (low to high near MJD 49950;
low to intermediate near MJD 50400; low to high near MJD 50850),
exhibit this behavior.

(d) During transitions, the $V-R$ color is nearly
constant: $\Delta$($V-R$)$<$0.03. 
We reiterate this point, because it may provide an
important clue to the physics of the transitions.

(e) The MACHO data set includes a long-lived intermediate state,
which likely started some time before the MACHO observations
commenced and which lasted during the first $\sim 700$ days
of monitoring. The MACHO data set also includes  a long-lived
high state, that started at roughly MJD 50875 and which lasted for
just over $600$ days.

(f) The MACHO data set does not include a comparably long low state;
the longest lasting low states have durations of $\sim 200$ days.

(g) For a  period of about $1200$ days, between the long-lived
intermediate state and the long-lived high state mentioned above,
the system experienced a sequence of shorter-lived ($200-250$ day)
states: low; high (punctuated by excursions to an intermediate state);
low; intermediate; low.

(h) The intermediate state that occurred during the $1200$ day interval
of more frequent transitions shows a steady declining trend.
Indeed, a general declining trend from high through intermediate and
to low luminosity was evident from $\sim$ MJD 50000
to MJD 50850. This overall trend is interrupted by the optical low states that
occurred immediately before and after the 1996 X-ray off state.

\subsection{X-ray Variations} 

Two X-ray states have been observed, an on-state, and an off-state.
Though measurements with different satellites/instruments are difficult
to compare, all evidence points towards a constant effective temperature
and luminosity, with absolute values (depending on the applied models) of 
$kT \sim 30-50$ eV, and $L \sim 2\times 10^{37} - 1\times10^{38}$ erg s$^{-1}$.
The ROSAT-detected X-ray off state represents a factor of $\gax 40$ variation 
in count rate, while the {\it Chandra}-detected X-ray off state represents
 a factor of $\gax 500$ variation in count rate
(based on a PIMMS predicted rate for CAL 83's normal X-ray on-state using
the ROSAT PSPC count rate; these predictions have been verified to be
true for 6 other supersoft X-ray sources observed with {\it Chandra} 
ACIS-S; Greiner \etal\ 2002).

An X-ray observation that occurred $\sim 21$ days prior to the 1996 
X-ray off-state showed the system was in its normal X-ray on-state
and was apparently neither becoming less luminous nor cooler.
An X-ray observation that occurred $\sim 100$ days after the 1996 
X-ray off-state again found the system to be in its  normal X-ray on-state.
These observations place strong constraints on the decay and growth time
of the X-ray flux and on changes in effective temperature,
both prior to and after the off-state.

\subsection{Correlation of optical and X-ray data}

(a) It is an interesting and unfortunate fact that no X-ray observations
were carried out during the long-lived intermediate and high optical states
covered by the MACHO data (for the preceding decades since 1980 optical 
observations were too sparse to allow a classification into one of the three
states).
The sole exception is the {\it Chandra} observation on MJD = 51513 which
occurred during the optical high state, but just before the system
made a sharp transition to the optical low state.

(b) The ROSAT, ASCA, and {\it Beppo}SAX
 observations occurred during the period of
most frequent optical-state transitions, between MJD 50025 and MJD 50625.
X-ray on states observed during this time occurred during optical low and
intermediate states.

(c) X-ray off states have occurred only during optical high states.
Note that the X-ray coverage is so limited that we cannot infer
that the X-ray emission is always detectable during low and intermediate
states. Nor do we know that the X-ray radiation is always undetectable
during optical high states.

(d) The very last MACHO observation shows a dramatic fading which
seems related to our Chandra-discovered off-state.
Fortunately, several optical observations occurred for about 30 days
after the Chandra observation (MJD = 51513) of the X-ray off-state, with
no indication of a decline.
This seems to indicate a clear delay between the X-ray off and optical low
states. When the optical decline occurs, however, it is relatively sudden.   
Indeed, the time between the last two MACHO observations is 5 days, showing 
that the optical fading by 0.7 mag happened in less than 5 days 
(between MJD 51542 and 51547), and 29--34 days after the Chandra observation.
Since the beginning (and end) of the X-ray off-state is unknown, the actual 
delay time between X-ray and optical low-state is longer than 29--34 days.

(e) 
The behaviour during the time of the April 1996 X-ray off-state 
(MJD around 50200) can be interpreted in the same way.
That is, the  X-ray  off state is followed by optical fading
around MJD = 50250 and continuing up to MJD = 50330).
If we adopt such an assignment, the delay of the optical fading
would be about 50 days after the
X-ray off-state, comparable to the interval observed in November 1999.
It is attractive to apply the same scenario to
explain the sequence observed for both X-ray off states.
Note that the Alcock {\it et al.} (1997) scenario 
of optical dipping during X-ray off-states
cannot apply to the {\it Chandra}-discovered off state.

\subsection{Possible Patterns of correlated optical/X-ray variation}

X-ray coverage of CAL 83 has been sporadic.
Even though the optical observations
by the MACHO team were regular, the average time between observations
was $8$ days, during which both X-ray and
optical fluxes were changing on comparable or possibly smaller time
scales. It is therefore entirely
possible that more frequent and regular observations  would find
behavior of a type we have not yet been able to observe. 

Two patterns of correlation are consistent with
the X-ray and optical data taken to date.

\begin{enumerate} 

\item X-ray off-states occur during optical 
high states, and X-ray on-states during optical low and intermediate state.

\item 
Optical variations (e.g. decline) follow about 30--50 days after an 
X-ray state transition (e.g. turn-off).
\end{enumerate}

\section{Understanding the Variations in the Context of Binary Models}

\subsection{X-Ray and Optical Variations in SSBs}

To determine the reasons for the X-ray and optical variations, 
we must consider the origin of radiation in each energy band. 
In the standard SSB models (van den Heuvel \etal\ 1992;
Rappaport, Di Stefano \& Smith 1994; Di Stefano \& Nelson 1996),
the copious soft X-ray radiation is generated by the nuclear burning
of matter near the surface of the white dwarf.
In order for a WD to burn matter as it accretes, the accretion  
rate must be 
near $\sim 10^{-7} M_\odot$ yr$^{-1}$. 

A key characteristic of SSBs, such as CAL 83 and \rxj0513\ is that, while 
the bulk of the X-radiation is emitted by the nuclear-burning white dwarf,
it is the accretion disk that is the primary source of radiation
at longer wavelengths (Popham \& Di Stefano 1996). 
The flux and spectrum of radiation
from the disk are largely determined by the reprocessing of energy
from the white dwarf and not directly by the accretion process.
The optical flux from the disk is significantly greater than   
that associated with white dwarf or its companion, the latter of which
likely explains the fact that the spectra of the companions 
in close-binary supersoft sources have not been detected.
Thus, although variations in the X-ray flux must be explained in terms of
variations associated with the white dwarf, optical variations
must be primarily associated with the disk.

\subsection{Variations of the X-Ray Flux} 

EUV radiation and soft X-rays are generated near the
white dwarf's surface, where nuclear burning is or has been occurring.
X-ray variability can be due to changes in the 
accretion rate, burning rate, or photospheric radius.
At rates larger than the burning rate, matter which accretes but fails to
burn may form a swollen envelope, or can be ejected in the form of winds.

\subsubsection{Does the WD Central Engine Turn On and Off?}

It is reasonable to ask whether the apparent cessation of X-ray radiation
at two times in the light curve of CAL 83
is likely to be a signal that nuclear burning had ceased, and that
the central source was no longer emitting X-rays. 
This is unlikely, because 
it takes time for the white dwarf to cool. Those novae which have
been observed as SSSs have taken months or even years to cool enough that
they are no longer observed as X-ray sources,
like GQ Mus (\"Ogelman \etal\ 1993); V1974 Cyg (Krautter \etal\ 1996), 
Nova LMC 1995 (Orio \& Greiner 1999).
We do not know the mass of the white dwarf likely
to be the central source in CAL 83.
However, the low X-ray luminosity (G\"ansicke \etal\ 1998)
as well as the white dwarf model fits to {\it Beppo}SAX data 
(Parmar \etal\ 1998) suggest that the mass is not high
($\sim$0.9--1.0 \msun).
A variety of computations  (e.g. Prialnik \& Kovetz 1995; 
Kato 1997, 1999) suggest that cooling  (i.e. turn-off)
times can be as short as 6--20 days only if the
mass of the white dwarf is near the Chandrasekhar limit; otherwise
the cooling time (for the emission to be shifted out of
the X-ray regime) is in excess of several months.
It therefore seems highly unlikely that
a cessation of  nuclear burning occurred around MJD $\sim$ 50200 in 1996,
leading to the X-ray off-state found
 by Kahabka \etal\ (1996). It is possible that 
nuclear burning ceased much earlier, but in this case, the X-ray detections 
prior to MJD $\sim$ 50200 should have shown some indication of evolution 
toward a cool state. Since there is no evidence of such evolution,
we conclude that the off-state in 1996 was unlikely to be related
to a complete cessation of nuclear burning and subsequent cooling
of the white dwarf. 

This does not mean there might not have been a modulation of
the bolometric luminosity during an interval of time
around the off state, perhaps due to a modulation of the 
accretion rate, or perhaps  due to a change in the short-term
average energy release associated with nuclear burning.
But it does mean that a hot white dwarf, with luminosity
roughly comparable to its value at earlier and
later times, was still present in
CAL 83, even during the X-ray off-state in 1996.
Occam's razor would therefore suggest that the WD was also hot during
the {\it Chandra}-discovered off-state. 

The question then becomes, if the hot white dwarf was there,
why did we not detect X-rays during these $2$ observations?
There are two possibilities.

\begin{enumerate}

\item The white dwarf's energy 
emission may be shifted toward longer wavelengths.
This could happen if, e.g., the radius of the photosphere increases.  

\item X-rays emitted by the white dwarf
could interact with matter in the vicinity of the system;
the energy thus absorbed would eventually be re-emitted at other wavelengths.

\end{enumerate}

\subsubsection{Photospheric Adjustments} 

Expansion of the photosphere can shift the peak of the spectrum toward
longer wavelengths.
Because the X-ray emission we detect from SSBs constitutes only
the high-energy tail of the emitted spectrum, even modest
shifts of the peak
to lower energies can shift the tail so that X-ray instruments
are no longer able to detect the source.
Modulations of the photospheric radius can be caused by
adjustments of the accretion rate, or by changes associated with
nuclear burning, even at relatively steady accretion rate.
Photospheric adjustments seem to be the  most direct way
to achieve the observed X-ray off-states, in cases in which the bolometric
luminosities do not apparently suffer a large decline and
an X-ray on-state re-occurs soon thereafter. 

Indeed this scenario has shown to work for the novae GQ Mus 
(\"Ogelman \etal\ 1993) and V1974 Cyg (Krautter \etal\ 1996), and has
been applied to similar sources like the SSB
RX J0513.9--6951 (Reinsch \etal\ 1996; Southwell \etal\ 1996)
or the symbiotic system AG Dra during outburst (Greiner \etal\ 1997).
\rxj0513\ is the most prominent example, varying quasi-periodically at 
both X-ray and optical wavelengths.
 Because the variations repeat at relatively short time intervals, 
the variations of \rxj0513\ are much better studied than those of
CAL 83. In particular, \rxj0513\  goes through
optical low-states lasting $\sim$30--40 days and repeating quasi-regularly
at 150--180 days intervals. 
The decrease in the optical flux of $\Delta$$V \sim 1$ mag
is accompanied by a reddening of $\Delta$($B-V$)$\sim$0.1--0.2 mag and
$\Delta$($V-R$)$\sim$0.1 mag (Reinsch \etal\ 1996). 
Soft X-ray emission is only observed during the short optical low-states, 
while during the long optical high states no X-ray emission is detected.
There have been two different suggestions to explain this behaviour:
\begin{enumerate}
\item If during optical high states the accreting white dwarf is slightly
 (by about a factor of 3) expanded and cool enough to not emit X-rays, then
a temporarily slightly reduced mass-transfer rate would cause the
photospheric radius to shrink. As a consequence, optical emission from the
white dwarf would be reduced, and later on the white dwarf would get hot enough
($kT \gax 20$ eV) to be detectable in X-rays (Pakull \etal\ 1993;
Southwell \etal\ 1996).
\item Taking into account the fact that in \rxj0513\ most of the light 
from the accretion 
disk is reprocessed emission, Reinsch \etal\ (1996) proposed that
it is not the white dwarf itself which is producing the varying optical 
emission, but the changes of the white dwarf induce
variations of the irradiation of the accretion disk 
which in turn are the cause for the observed optical variations.
\end{enumerate}

Interestingly enough, the physical parameters
of \rxj0513\ (X-ray flux, temperature, orbital period) are
similar to those of CAL 83. Consequently, both models
 have been invoked to explain CAL 83's variations (Kahabka 1998).
Alternatively, Alcock \etal\ (1997) suggested a cessation of nuclear burning
on the white dwarf in CAL 83, leading to a behaviour  similar to the 
decline phase of a nova. 
We have argued above that a cessation of nuclear burning in CAL 83 is 
very unlikely. We will argue below that the Kahabka model 
with its additional assumption of a particularly shaped mass transfer
modulation for CAL 83 does not
explain all the presently available data. Instead, a more 
complicated variant of model (2) seems to be required:
\begin{enumerate}
\item The variability pattern is much more complicated in CAL 83
than in \rxj0513, i.e. not just a relatively fast switching between 
two brightness levels.
\item 
CAL 83 shows no color changes (Figs. 1, 2). 
Both photospheric adjustments as well as a varying wind (absorption) 
would predict substantial color changes ($\gax 0.1$ mag), 
and the binary inclination is comparable in both systems.
Though CAL 83 and \rxj0513\ have slightly different disk sizes
as implied by their different orbital periods (1.04 days vs. 0.76 days)
and luminosities, the relative changes (in \mdot, color etc.) should
be similar.
\item
Sophisticated modelling of the UV/X-ray observations of CAL 83 has resulted
in a bolometric luminosity of (0.7--2)$\times$10$^{37}$ erg s$^{-1}$ 
(G\"ansicke \etal\ 1998). The inferred mean luminosity averaged over
the last $\sim$10$^5$ years as derived from the \ion{[O}{III]} and H$\alpha$ 
lines of the ionization nebula is (1--7)$\times$10$^{37}$ erg s$^{-1}$ 
(Remillard \etal\ 1995), implying that X-ray off states should
represent only a small fraction over the last $\sim$10$^5$ years.
However, relating X-ray off states to optical high states
(above pattern 1) would be statistically inconsistent with such
a small fraction, since CAL 83 was in an optical high state for $\sim$30\%
over the last 10 years.
\item
The overshooting after transitions from optical low or intermediate state to
high state is difficult to explain by  photospheric radius changes.
\end{enumerate}
We also note that even the expanded white dwarf cannot outshine the
accretion disk, unless it grows to 5$\times$10$^{11}$ cm, larger than
the binary separation of 3.7$\times$10$^{11}$ cm.

\subsubsection{Absorption} 
 
Absorption due to material above the burning surface
could also have the effect of shifting the effective radius.
If, for example, (1) more matter is accreting per second than can,
on average,  be burned
during one second, and (2) not all of this matter can be
instantaneously ejected, it does not take long for the
``extra" mass to interfere with the detection of X-rays. 
Suppose that the ejected winds are preferentially emitted in a cone (jet)
oriented perpendicular to the accretion disk
as deduced from the broad He and H$\alpha$ wings (Crampton \etal\ 1987;
Cowley \etal\ 1998). During an interval in
which a substantial fraction of material incident from the donor
is neither burned nor ejected as winds, $\sim 10^{19}$ g s$^{-1}$
may accumulate over an area smaller than $(10^9)^2$ cm$^2$.
X-ray radiation could be quenched on relatively short time scales,
as unburned matter begins to seep 
upward through what had been the wind cone.
The seeping matter can affect both our view of the disk and also
the radiation incident on the disk from the central star.
Note that, in the case of CAL 83, the blocking of the disk from our 
view is unlikely to be the only cause of optical variation. Although
this model  is consistent with the lack of color change, it
would not lead to a spike in the optical emission when the
disk again becomes visible. The existence of the overshooting apparent
in the light curve tells us that some real adjustment of
the disk is likely to be taking place as it moves from one equilibrium
state to another.     

\subsection{Variations of the Optical Flux}

Variations of the optical flux are most likely to be associated
with changes in the accretion disk. 
The accretion disks surrounding the \WD\ in SSBs are bright,
generally more than an order of magnitude brighter than 
the accretion luminosity (viscous heating).   
This is because, in contrast to the situation
for canonical cataclysmic variables or neutron star binaries, 
nuclear burning of accreted material releases 
($\sim 20$ times) more energy, than the potential
energy lost and dissipated through accretion.
Reprocessing of this radiation by the disk is of particular
importance in CBSSs (e.g. Popham \& Di Stefano 1996; Matsumoto \& Fukue 1998;
Fukue \& Hachiya 1999; Suleimanov \etal\ 1999).
If $\sim 20\%$ of the energy emitted by the white dwarf
is reprocessed by the disk,
the disk luminosity will be $10^{35}-10^{37}$ erg s$^{-1}$.
Farther from the \WD, the disk is cooler due to the less pronounced heating, 
becoming the dominant
source of UV and optical light. The height of the outer disk flares
(Popham \& DiStefano 1996; Meyer-Hofmeister \etal\ 1997). 
The geometry of the disk emerges from first-principles 
calculations (Popham \& Di Stefano 1996) and, for the eclipsing
CBSS CAL 87, can also be derived phenomenologically (Schandl \etal\ 1997).

If roughly $20\%$ of the energy released due to nuclear
burning is reprocessed by the disk, and if roughly
one half of the accretion energy is dissipated in and radiated from the disk,
then accretion energy accounts for only 
$\sim 10\%$ of the energy emitted by the disk. 
Changes in the accretion rate produce changes in the disk
luminosity on the order of a few (perhaps as much as $10$) percent.
These changes are likely to occur prior
to any changes in X-ray flux due to a changing
rate of fuel delivery by the disk.

A cessation of nuclear burning could eventually (after cooling)
cause the disk to become 2 orders of magnitude less luminous, even if 
accretion continued at a comparable rate. 

Unless they constitute just a small fraction
of an upward or downward trend, luminosity changes by a factor of $\sim 2$
are most likely due to changes directly related to the
reprocessing of radiation from the \WD. 
Reprocessing changes are of two types. First, reprocessing
that occurs as radiation directly from the central star hits and
interacts with the disk, is naturally influenced by changes in the spectrum
or flux of radiation from the central star. 
Second, heating from the central star through the disk will also be
affected.  
The  flow of heat
is influenced by the size and temperature of the photosphere as well as
by the instantaneous luminosity of the \WD.  This heating is 
communicated  from the \WD\ through matter in the disk. The disk response
can therefore lag the X-ray response.

\subsection{Correlations Between X-Ray and Optical Flux} 

The primary cause for the variations is either a modulation
of the rate at which the companion donates matter, or a
modulation of the rate of nuclear burning
(though the observed time scale of state changes in CAL 83 is very 
short for a supposed white dwarf mass of $\sim$1 \msun).
In each case, the bolometric luminosity will change, even as the changes
affect different parts of the spectrum differently.
Prompt anticorrelations between X-ray and optical flux
suggest photospheric changes that shift the flux distribution
 from the X-ray range  towards longer wavelengths
and immediately influence the disk by affecting the spectrum
of light incident on it from the white dwarf. Delayed correlations
or anticorrelations suggest that matter flowing to the white dwarf
may be responding to different local conditions as disturbances due to
changing accretion rates and differences in heating from the
white dwarf propagate through the disk.

\section{Summary and Prospects}
CAL 83 is one of the flagship sources of the class of supersoft X-ray binaries.
It may therefore seem strange that CAL 83 was discovered to have an
X-ray off state $16$ years after it was first observed by {\it Einstein}.
In common with many other X-ray sources, CAL 83
has been observed only sporadically. It may therefore
not be so surprising that we have little information about
its X-ray variations. Furthermore, although RXTE has conducted regular all-sky
monitoring, its detectors are not sensitive to the soft
radiation emitted by CAL 83.

Two alternative interpretations of the pattern of X-ray and optical
variations are consistent with the data on CAL 83: 
(1) X-ray off-states occur during optical 
high states, and X-ray on-states during optical low and intermediate state, or
(2) optical variations follow about 30--50 days after an X-ray state 
transition.

We have shown above that neither the simple picture of photospheric
radius changes nor pure absorption effects can be the primary cause for
these observed patterns in CAL 83.
Instead, changes in the photospheric radius lead to
changes in the irradiation and heating of the disk, perhaps
at a time when the accretion rate is also (or has been) changing.
It is this suite of changes, accompanying any photospheric change,
that modulate the optical luminosity. Optical variations can
therefore have a more complex
relationship to photospheric expansion or contraction.

One of the things we have learned, however, is that
optical monitoring programs,
like those carried out to search for evidence of
microlensing, can complement the information gleaned from 
X-ray observations. 
For \rxj0513, for example,  a combination of early optical
observations specifically designed to learn more about 
the X-ray source (Reinsch \etal\ 1996)
and subsequent monitoring by the  MACHO team (Alcock \etal\ 1997),
allow us to now use the optical data to predict the 
X-ray state. This also seems to be true of the 
VY Scl stars, in which a proposed anticorrelation 
between optical and X-ray fluxes allowed us to discover that
at least one member of this class is a supersoft X-ray binary
(Greiner \etal\ 1999). For CAL 83, we clearly need a better record of its
X-ray light curve, coupled with optical
monitoring, to better understand the relationship between
its optical and X-ray flux.

Based on the pattern observed to date and the previous 
discussion it is clear that 
simultaneous and frequent multiwavelength observations of CAL 83 are needed
to test physical models for the source and its variability.
Such observations can provide important clues to
the physics of this interesting system and the SSB class
to which it belongs.

\acknowledgements
We are highly indebted to A. Becker for making the calibrated MACHO data
available to us. We also want to thank Andrea Dupree for 
enlightening discussions.
Thanks also to the CfA data processing group for extensive data
validation checks of the Chandra observation of CAL 83 in Nov. 1999.
We appreciate the generosity of Eric Gotthelf for making available his
re-analysis of the CAL Einstein data for publication herein (see Appendix).
We are grateful to the referee, P. Kahabka, for the
careful reading and detailed comments which improved the presentation
of the data.
This work was supported in part by NSF under INT-9815655 and by NASA under
NAG5-10705.

\bigskip\bigskip\bigskip

\newpage

{\it Appendix.}

\smallskip

Below are the results of a re-analysis of the Einstein IPC data
by Eric Gotthelf (Columbia University, USA) from 1996, which have
been mentioned as ``priv. comm.'' in Kahakba (1998) but were never
published. These results correct the wrong count rates of Brown \etal\ (1994).
As can be seen from the table below, there does not seem to be any
significant source variability ($<$10\%). Notice that the observation closest
to the optical axis has the lowest count rate. In contrast, one would expect
a systematic increase in the count rate if the difference was due
to instrumental effects.

\begin{table*}[hb]
\caption{{\it Einstein} IPC observations of CAL 83 (by E.V. Gotthelf)}
\vspace{-0.1cm}
\begin{tabular}{cccrrccl}
  \hline
   \noalign{\smallskip}
  Seq\# &   ObsTime &   Pointing (B1950) & Roll~ & Livetime & Src Cts$^{(a)}$ &
   Offaxis & Rate (error)$^{(b)}$ \\
        &    (JD)   &     R.A. ~~~  Dec. & (deg) & (sec)~~ &  (2-10 ch) &
    (arcmin) & ~~~(r=3\amin) \\
  \noalign{\smallskip}
   \hline
   \noalign{\smallskip}
 2417 & 2443974.09981 & 5 43 43 --68 11 59 & 66~ &  573~ & 61  & 11 & 0.090(0.018) \\
 2418 & 2443972.99426 & 5 49 04 --68 11 59 & 68~ & 1690~ & 16  & 31 & 0.060(0.05) $^{(c)}$ \\
 2430 & 2443972.78994 & 5 47 43 --68 41 59 & 68~ &  899~ & 80  & 28 & 0.120(0.02) \\
 6301 & 2444280.04540 & 5 43 12 --68 58 19 & 124~& 23361~& 1419& 35 & 0.108(0.003) \\
 7109 & 2444279.08445 & 5 43 29 --67 50 59 & 125~& 13841~& 1049& 32 & 0.120(0.004) \\
  \noalign{\smallskip}
   \hline
  \noalign{\smallskip}
\end{tabular}

\noindent{{\it Notes:}

 ($^a$) Background subtracted source counts in a 3\amin\ radius aperture 
  centered on CAL 83. \\
 ($^b$) Summed exposure corrected count rate in a 3\amin\ radius aperture 
  (error $\sqrt{(S+B)/(S-B)}$ ), where $S$ and $B$ are the source and 
  background count rate, respectively.  \\
 ($^c$) This number is unreliable. The source was partially obscured by an 
  IPC rib.}
\end{table*}

\end{document}